\documentclass[
    ,final            % use final for the camera ready runs
  ]
  {aipproc}

\usepackage{amsmath}
\usepackage{subfigure}
\usepackage{cellspace}
\usepackage{multirow,bigstrut}
\usepackage{tabularx}

\layoutstyle{6x9}

\usepackage{url}
\usepackage{bm}

\usepackage{booktabs}           % Buchtabellen
\usepackage{placeins}		% \FloatBarrier

\begin{document}

\title{Laser Ablation of Al-Ni Alloys and Al-Ni Layer Systems simulated with
Molecular Dynamics and the Two-Temperature Model}


\author{Dennis-Michael Rapp$^\dagger$}{
  address={
    Institut f\"ur
    Materialpr\"ufung, 
    Werkstoffkunde und Festigkeitslehre, Universit\"at Stuttgart}
}

\author{Alexander Kiselev}{
  address={
  Institut f\"ur Funktionelle Materie und
    Quantentechnologien, Universit\"at Stuttgart}
}

\author{Hans-Rainer-Trebin}{
  address={
  Institut f\"ur Funktionelle Materie und
    Quantentechnologien, Universit\"at Stuttgart}
}

\author{Johannes Roth}{
  address={
  Institut f\"ur Funktionelle Materie und
    Quantentechnologien, Universit\"at Stuttgart}
}

\begin{abstract}
  Laser ablation of Al-Ni alloys and Al films on Ni substrates has been
  studied by molecular dynamics simulations (MD).
  The MD method was combined with a two-temperature model to describe
  the interaction between the laser beam, the electrons and the atoms.

  The challenge for alloys and mixtures is to find the electronic parameters:
  electron heat conductivity, electron heat capacity and electron-phonon
  coupling parameter. The challenge for layered systems is to run simulations
  of an inhomogeneous system which requires modification of the simulation
  code. 
  
  Ablation and laser-induced melting was studied for several Al-Ni
  compounds. At low fluences above the
  threshold ordinary ablation behavior occurred while at high fluences the
  ablation mechanism changed in Al$_3$Ni and AlNi$_3$ from phase explosion to
  vaporization.

  Al films of various thicknesses on a Ni substrate have also been
  simulated. Above threshold, 8 nm Al films are ablated as a 
  whole while 24 nm Al films are only partially removed. Below threshold,
  alloying with a mixture gradient has been observed in the thin layer system.
  
\end{abstract}

\maketitle

\section{Introduction}

Molecular dynamics simulations (MD) of laser ablation with femto-second pulses
is well established nowadays
\cite{Schaefer2002,itapdb:sonntag2009,itapdb:Roth2011}. To describe 
the interaction of the laser and the metal accurately, the molecular dynamics
(MD) simulation method has to be enhanced by the two-temperature model (TTM)
where  electrons and the ions (or the lattice) are described by different
temperatures \cite{itapdb:Anisimov1974}.

Most studies of laser ablation to date deal only with pure
metals. To our knowledge, there is only one publication on complex alloys
\cite{itapdb:sonntag2009a}, and a few on Au-Cu and Ag-Cu layer systems
\cite{aucu-ablat,agcu-run,exalcu}. See also the femtosecond melting of Au-Cr
layer systems \cite{aucr1,aucr2}.

There are on the other hand very interesting studies of shock waves in AlNi
multilayer materials
\cite{alni-shocksim,alni-laminate,alni-shocksim3,alni-shocksim4,alni-amorph,alni-react,alni-last}.  
They reveal exothermic reactions and alloying of the multilayer systems under 
shock wave treatment. Similar phenomena are expected to occur in the AlNi
system under laser treatment.

Furthermore, nickel aluminides are technologically very interesting
materials which may have better properties than steel for example. The
melting points are higher than for pure aluminum, but they are lower than
for pure nickel. The intermetallic phases possess lattice structures which
reduce 
the mobility of dislocations and defects and reduce diffusivity. This leads to
superior mechanical properties at higher temperatures. The yield strength
increases with temperature in contrast to steel. Increased thermal heat
conductivity and corrosion resistivity renders them 
candidates for high temperature applications in airplanes and gas turbines. 
But these applications also require precise processing of the
materials. This is were ablation with femto-second lasers comes into
play.  

Details of this work have been published already \cite{rothcola}. Therefore the
method and results will only shortly be summarized. The influence of the
materials 
anisotropy will be discussed and new results and analysis of Al films on Ni 
substrate will be reported. 

\section{The simulation model}

The basic equations of the two-temperature model are generalized heat
conduction equation for the electrons and the atoms separately: 
\begin{eqnarray}
  C_{e}(T_{e}) \frac{\partial T_{e}}{\partial t} & = & \nabla [K_{e} \nabla
  T_{e}] - 
  \kappa (T_{e} - T_{a}) + S(\textbf{x},t) \label{eq:TTM1}\\
  C_{a}(T_{a}) \frac{\partial T_{a}}{\partial t}  & = & \nabla [K_{a} \nabla
  T_{a}] + 
  \kappa (T_{e} - T_{a}) \label{eq:TTM2}
\end{eqnarray}
Equations (\ref{eq:TTM1}) and (\ref{eq:TTM2}) describe the time
evolution of the electronic ($T_{e}$) and the atomic ($T_{a}$) or lattice
temperature within the material. $S(\textbf{x},t)$ is the external laser
field. $C_{e,a}$ are the heat capacities, $K_{e,a}$
the heat conductivities, $\kappa$ the electron-phonon coupling constant.
In general these physical properties are functions of $T_e$ and $T_a$.
With these equations (\ref{eq:TTM1}) and (\ref{eq:TTM2}) the laser field
  is coupled physically meaningfully 
  into the system: first the energy is brought into the electronic system via a
source term $S(\textbf{x},t)$. Then the electronic system transports the heat
into the bulk and at the same time interacts with the atoms.

To work on an atomic scale equation (\ref{eq:TTM2}) has to be replaced by
molecular dynamics. Instead 
of (\ref{eq:TTM2}) the following equations of motion are solved together with
(\ref{eq:TTM1}) for each atom $j$:
\begin{equation}
   m_{j} \frac{d^{2}\textbf{x}_{j}}{dt^{2}} = - \nabla_{\small \textbf{x}_{j}}
  U(\lbrace \textbf{x}_{k} \rbrace) - \frac{\kappa}{C_{a}} \frac{(T_{a}
  -T_{e})}{T_{a}} m_{j} \frac{d \textbf{x}_{j}}{d t}.
\end{equation}
The velocity-dependent friction term represents the coupling
between TTM and MD.  

The differential equation (\ref{eq:TTM1}) for the electrons is solved on a
lattice by a finite difference scheme (FD). For pure metals and alloys all FD
cells are equivalent.

\subsection{Generalization of the model}

The electron heat conductivity parameter $K_e$ is a scalar function as long as
the material is isotropic, cubic or nearly cubic which is the case for the AlNi
system. In general equation (\ref{eq:TTM1}) has to be generalized to a
tensorial function. Fortunately Al$_{13}$Co$_4$ can be treated as
orthorhombic, thus only diagonal terms are required. But since Al$_{13}$Co$_4$
is a layered material it strongly deviates from isotropy. The other electronic
parameters remain unchanged since no spatial gradients of the heat capacity
and electron-phonon coupling are present in equation (\ref{eq:TTM1}).

In the simulations of layered systems FD cells will be present 
with varying and mixed occupancy. The question is then how to find the correct
averages for the electronic parameters.

The gradient term of the temperature in Eq.~\ref{eq:TTM1} can be replaced by
the second derivative as long as the electron heat conductivity $K_e$ is a
constant. If $K_e$ 
becomes position dependent it is computed as the harmonic average of two
neighboring FD cells at $x-1$ and $x$ as recommended in \cite{bunge-skript}
\begin{equation}
  K_e=\frac{K_e^{x-1}K_e^x}{K_e^{x-1}+K_e^x}.
\end{equation}
This definition works even if the heat conductivities deviate strongly. Thus
it is possible to simulate the electron part of multi-component samples.

Usually, the inverse absorption length $\mu$ of the laser radiation is regarded
as a constant and the absorption is modeled by the standard
Lambert-Beer-law. This approach is no longer valid for the layered
systems.
Therefore, the absorption length is replaced by an effective length.
For example, a layer which absorbs twice as much energy as
the reference layer is taken into account with twice its thickness.
The energy absorbed in a certain cell is then computed by comparing the
difference of the laser intensity to the left and right:
\begin{equation}
  \Delta E = \Delta t_{FD} (S(d_l,x)-S(d_r,x))
\end{equation}
with the absorbed energy $\Delta E$, FD-timestep $\Delta t_{FD}$, laser
intensity $S$, position $x$ and effective depths $d_r$ and $d_l$.
The laser absorption can still be modeled by an exponential law
$\exp(\mu_\mathrm{eff}d)$ with an effective inverse absorption length
$\mu_\mathrm{eff}$.
The dependency of the injected energy on the decrement of the laser power
guarantees that the applied fluence $S_0$ is deposited in the sample.

\subsection{Simulation code and interactions} \label{portfitdis}
All simulations have been carried
out with \texttt{IMD}, the ITAP Molecular Dynamics 
package which is described in detail in
\cite{itapdb:Stadler1997a,itapdb:Roth2000} and for 
laser ablation in \cite{itapdb:Roth2011}. The program is available from
\texttt{github} \cite{imd}.

The atomistic interactions were modeled by EAM potentials. For the AlNi
system a set of potentials was computed by Purja Pun and Mishin
\cite{mishin2009-NiPot}. For Al$_{13}$Co$_4$ the potentials \cite{alco} were
computed by fitting effective potentials to ab-initio simulations with
\texttt{potfit} \cite{potfit}. 

%\subsection{Stability considerations}
%For constant TTM parameters the stability criterion relating MD timesteps
%$\Delta t_{MD}$  and FD timesteps $\Delta t_{FD}$ is given by
%\cite{Schaefer2002}: 
%\begin{equation}
%  N=\frac{\Delta t_{MD}}{\Delta t_{FD}}= 2\Delta t_{MD} \frac{K_e}{h C_e}
%\end{equation}
%where $h$ is the smallest dimension of the FD cells. This equation is replaced
%by 
%\begin{equation}
%  N=\frac{\Delta t_{MD}}{\Delta t_{FD}}= 2\Delta t_{MD} \frac{\max K_e}{h \min
%    C_e}
%\end{equation}
%where $\max K_e$ is the maximal heat capacity and $\min C_e$ the minimal
%heat capacity occurring in the simulations. The largest $K_e$ is limited to the
%maximal $K_e$ of a material component due to the harmonic average applied for
%the heat conductivities. Very low values of $C_e$ can occur for partially
%occupied cells if density-based averages are applied for the heat capacities. 
%The number of steps to fulfill the stability criterion is then much higher
%than in the case of volume-based averages where the minimal $C_e$ cannot be
%lower than the smallest $C_e$ of a component since it is computed from the
%weighted average of the material components in a cell.

%%%%%%%%%%%%%%%%%%%%%%%%%%%%%%%%%%%%%%%%%%%%%%%%%%%%%%%%%%%%%%%%%%%%%%%%%%%%%%%%

\section{Overview of laser ablation of Al-Ni alloys}

Details of the laser ablation and laser induced melting have been presented in
\cite{rothcola}. The pure phases and alloys Al, AlNi and AlNi$_3$ are
cubic. Thus the isotropic equations apply. Only Al$_3$Ni is orthorhombic but
it weakly deviates from isotropy. Thus it is also treated as
an isotropic material. Fortunately, all required parameters could be obtained
from literature except for the electron-phonon coupling constant which was
calculated with the formula of Wang et al. \cite{wang1994-ElPhHerleitung}:
\begin{equation}
\kappa = \frac{\pi^4 \left( k_B v_s n_e \right)^2}{18 K_e(T_e)}
\end{equation}
which gives the relation of $\kappa$ to the velocity of sound $v_s$, the
electron density $n_e$, and the electronic heat conductivity $K_e(T_e)$.

\subsection{Results for the pure metals Al and Ni}

The pure metals have been used as a benchmark for the generalized simulation
program. The laser ablation and laser-induced melting results compare well
with data from different sources, experiment and simulation.

\subsection{Results for the Al-Ni alloys}

Laser ablation and laser-induced melting has been studied for the alloys
Al$_3$Ni, AlNi, and AlNi$_3$. Other intermetallic phases exist in the Al-Ni
system but these are more complicated. The laser induced melting behavior has
been simulated for several sample sizes. The melting depths increases as
predicted by the well-known model of Preuss et al. \cite{preuss1995-exp} for
fluences up to at least 1000 J/m$^2$, but then levels off for the short
samples. The ablation depth shows strong fluctuations but can also be modeled
by the Preuss formula. At low fluences phase explosion is observed. For
Al$_3$Ni the mechanism changes to vaporization at about 3000 J/m$^2$. The same
is true for AlNi$_3$ at 900 J/m$^2$. The behavior of this phase is more
erratic, however, since it is accompanied by a phase transition from the
orthorhombic phase to fcc. Only for AlNi$_3$ it was possible to find some
experimental data \cite{pollock2007-AlNi3Abl} but even these data had to be
reanalyzed. The experimental data are rather close to the high fluence
behavior observed in the simulations. A further discussion of the results of
the other compositions is postponed due to the lack of comparable data. 

%%%%%%%%%%%%%%%%%%%%%%%%%%%%%%%%%%%%%%%%%%%%%%%%%%%%%%%%%%%%%%%%%%%%%%%%%%%%%%%%

\section{Anisotropic materials}\label{sec:qk}

All the structures simulated in the Al-Ni-system where more or less
isotropic. Thus no directional-dependent behavior was expected.

Previously we also studied a strongly anisotropic material, the approximant
Al$_{13}$Co$_{4}$\cite{itapdb:sonntag2009a} of the decagonal quasicrystal
phase Al-Ni-Co. The structure 
of this alloy features properties of periodic and quasiperiodic solids: it
consists of quasiperiodic planes stacked periodically with a period of 14 \AA~.
Therefore the heat conductivity $K_e$ possesses
a directional dependency, while the smaller effect on the
electron-phonon coupling is neglected. Due to the great interest in
quasicrystals it was possible to find the electronic parameters for this
structure. Interactions were computed with \texttt{potfit} as described in
the section about the interactions.

The melting depth turned out to be less than half the size of aluminum at the
same fluence. The melting depth was at most 20\% higher perpendicular to the
layers than in the layers, although the heat conductivity perpendicular to the
layers was  about 2.2 times larger than in the layers.
An explanation for the weak direction dependence is given by the relatively
high electron-phonon coupling of Al-Co (five times larger than that of
aluminum). This leads to a very short electron-lattice relaxation time
($\tau_{el-ion}\approx C_e/\kappa$) which shortens the diffusive heat
conduction of the electrons considerably.

%%%%%%%%%%%%%%%%%%%%%%%%%%%%%%%%%%%%%%%%%%%%%%%%%%%%%%%%%%%%%%%%%%%%%%%%%%%%%%%%

\section{Laser ablation of Al films on Ni substrate}

\begin{figure*}[!htb]
  \parbox{\textwidth}{
  \includegraphics[width=0.99\textwidth]{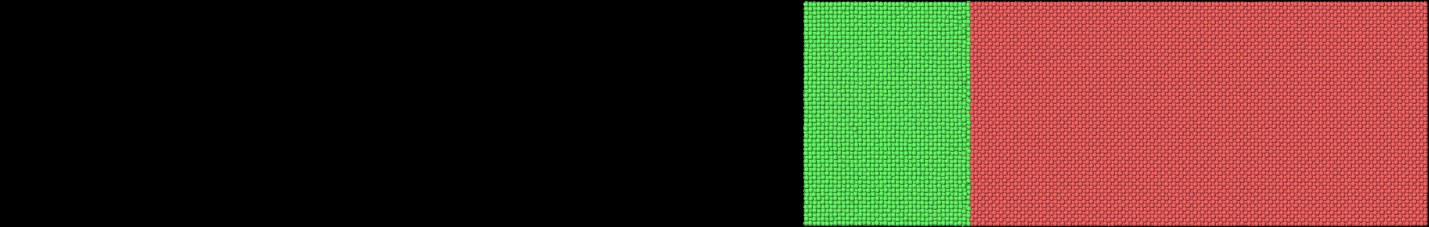}\\[0.1em]
  \includegraphics[width=0.99\textwidth]{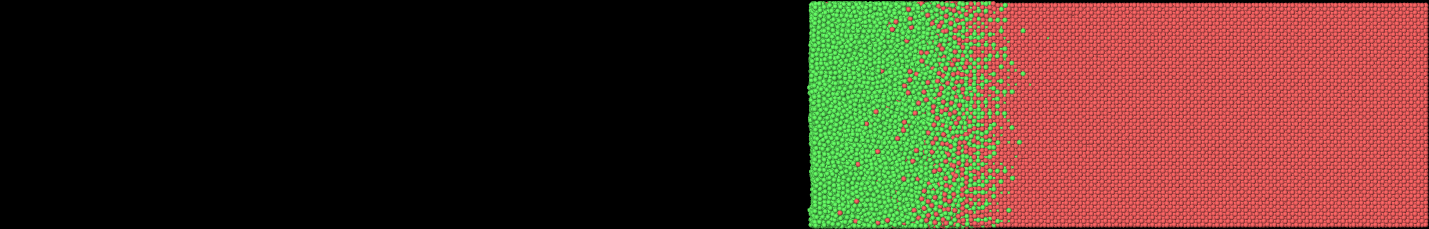}\\[0.1em]
  \includegraphics[width=0.99\textwidth]{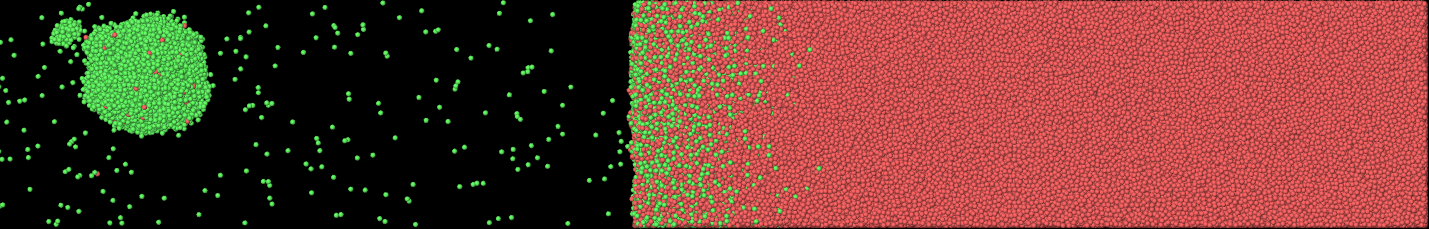}
  }
  \caption{Top: initial state of the thin layer sample before irradiation,
    center: irradiation below threshold, bottom: irradiation above
    threshold. Green: Al, red: Ni. Picture prepared with OVITO \cite{ovito}.} 
    \label{multilayer}
\end{figure*}

Two cases of an Ni substrate covered with Al were studied. The thin 
layer sample is covered by a 8 nm Al film, the thick layer sample by 24 nm Al.
A sketch of the observed mechanisms is
given in Fig.~\ref{multilayer}. In the figure at the center alloying was found
below threshold, at the bottom the roll off of most of the Al-layer was
observed above threshold with a sheet of an AlNi mixture staying behind. 

For the layer simulations samples have been prepared with about two
million atoms and a volume of 230 $\times$ 11 $\times$ 11 nm$^3$. 

\subsection{Results for the thin Al-layer}

Since Al has a much lower melting point than Ni it will melt already at
a low fluence of less than 40 J/m$^2$. Together with the lower heat
conductivity of Ni this leads to a plateau in the melting depth {\it vs.}
fluence at 8 nm until Ni starts to melt at 250 J/m$^2$. In general the
behavior is as given by the model of Preuss et al. \cite{preuss1995-exp}.   

The ablation fluence of the Al film is 500 J/m$^2$ at a depth of 10 nm, which
indicates that the Al film is ablated as a whole. Below the ablation threshold
voids are formed in the Al film and phase explosion occurs above threshold. 

\subsection{Results for the thick Al-layer}

The exponential decay of the absorbed laser fluence causes the absorption of
the major part of the energy within the Al-layer. The melting of the
Ni-substrate occurs preferably by propagation of the energy through heat
conduction. 

The melting depth again shows a plateau at about 24 nm if plotted {\it vs.}
fluence. This is caused by the complete melting of the Al-layer at fluences of
about  
60$\pm$ 20 J/m$^2$ until Ni starts to melt at a fluence of 360 J/m$^2$. The
general behavior is as predicted by the 
model of Preuss. The reason for the slightly higher damage threshold as
compared to the thin layer is the higher heat conductivity of the thick layer
which removes energy from the top of the sample. This behavior has been
confirmed in experiment \cite{guedde1998-NiDamThs}.

Below the ablation threshold bubbles are formed in the Al-film.
Ablation now occurs at 460 J/m$^2$. The ablation depth indicates that the
Al-layer is again ablated completely. But the behavior changes above about 900
J/m$^2$. The ablation depth decreases. The reason is that the Al is vaporized
now, but part of the material precipitates on top of the liquid
Ni-substrate. A mixed zone is generated where nickel aluminides can form.  

Thomas et al.\ \cite{aucu-ablat} have studied 30 nm gold films on copper
substrate. This system is similar to our thick Al layer. They observe
sub-surface melting of the copper which they attribute to the higher
electron-phonon coupling of gold. In the Al-Ni system such an effect was not
observed. The reason are the melting temperatures. The electron-phonon
coupling of Al is also larger than that of Ni, but the melting temperatures of
Au and Cu differ by about 20\% while the melting temperature of Al is about
half the value of Ni measured in Kelvin. 

\subsection{Formation of nickel aluminides}

\begin{figure*}[!htb]
  \parbox{\textwidth}{
  \parbox{\textwidth}{
    \includegraphics[width=0.32\textwidth]{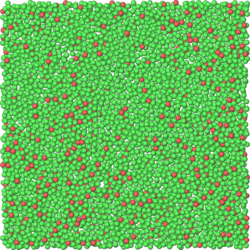}
    \includegraphics[width=0.32\textwidth]{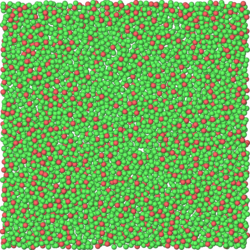}
    \includegraphics[width=0.32\textwidth]{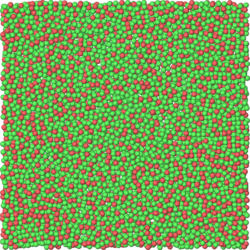}
  }
  \parbox{\textwidth}{
    \includegraphics[width=0.32\textwidth]{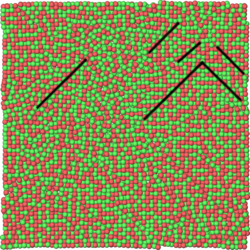} 
    \includegraphics[width=0.32\textwidth]{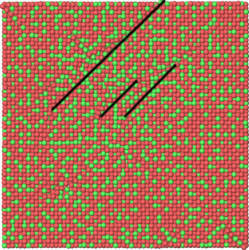} 
    \includegraphics[width=0.32\textwidth]{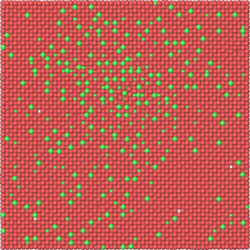}
  }
  }
  \caption{Slices through the sample at depths 5.5, 6.5, 7.5, 8.5, 9.5, 10.5
    nm (left to right, top to bottom). A few defects are indicated in the 8.5
    and 9.5 nm slices. Picture prepared with OVITO
    \cite{ovito}.}
  \label{slices}
\end{figure*}

The formation of nickel aluminides in the thin layer sample is shown in
Fig.~\ref{slices} at depths from
5.5 to 10.5 nm. Obviously the atom fractions are not constant but vary with
depth. The reason is diffusion. It has to be kept in mind that Al is still
molten down to 7.5 nm, while Ni has solidified largely. Starting at 8.5 nm a
fairly regular  structure is seen which indicates the formation of
non-stoichiometric Al-Ni mixed crystals. Deeper into the sample the Ni
fraction increases and the formation of AlNi$_3$ crystals and the substitution
of Ni by Al occurs. Diffusion will further generate all kinds of AlNi alloys
with varying concentration. Obviously this will not be the equilibrium crystal
structures. 

\begin{figure}[!htb]
  \includegraphics[width=0.49\textwidth]{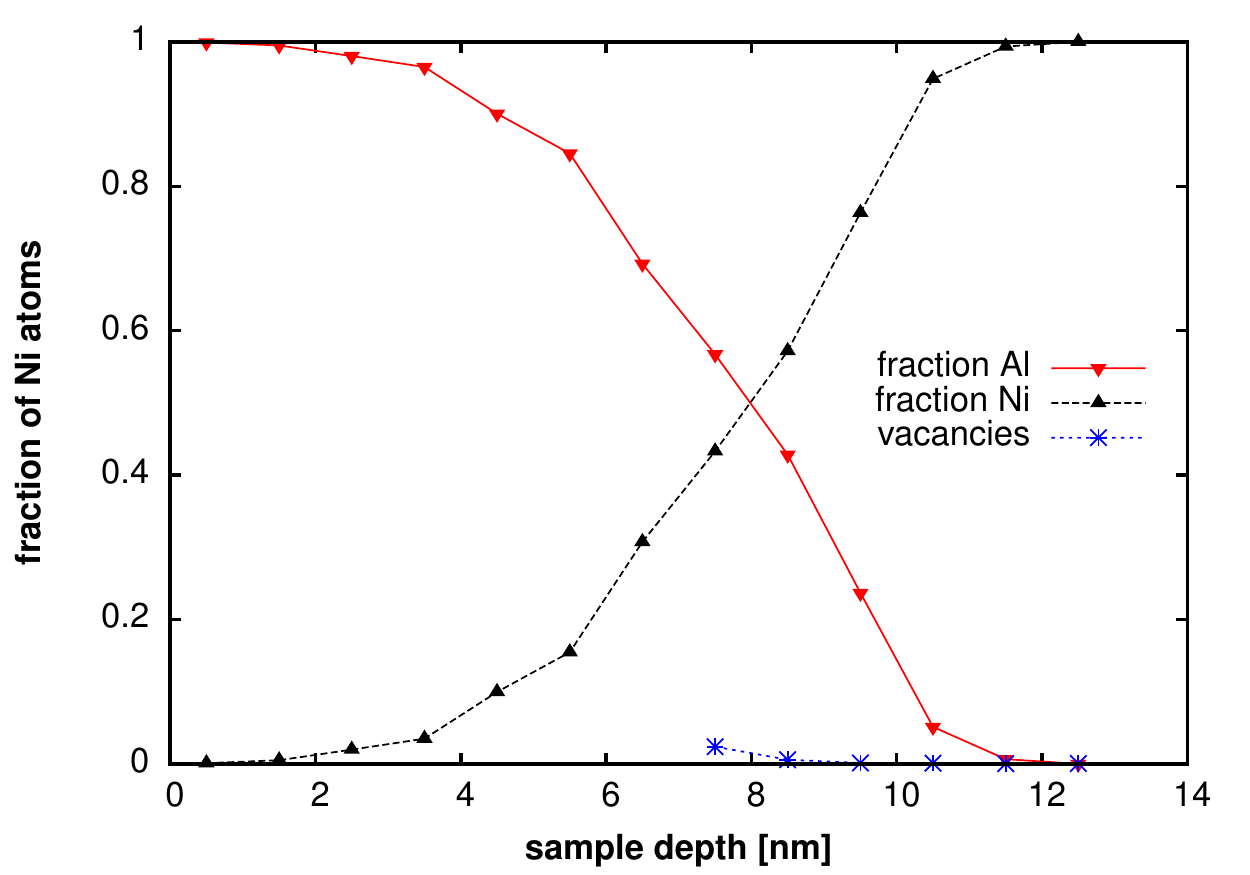} 
  \caption{Transition of the composition from the sample surface to the
    bulk. The original surface was at 8 nm, after irradiation the interface
    between liquid and solid lies at about 7 nm. Vacancies cannot be defined
  in the liquid part.}
  \label{transit}
\end{figure}

Fig.~\ref{transit} shows the transition from the Al film to the Ni substrate
after irradiation. The transition is asymmetric, since the Al film is liquid
while the Ni substrate is recrystallized. Thus the mobility of the minority
atoms is different in addition to different diffusivities. The number of
vacancies in the solid is measurable, but is at most 2\%. In the liquid
vacancies cannot be defined. In contrast to the observations of Cherukara et
al.\ \cite{alni-last} in shock simulations no preference for the
formation of Al$_3$Ni was found. The density in the liquid phase is that of
liquid Al and increases linearly starting at 4 nm until it reaches the density
of solid Ni at 10nm. The transition region is about twice as wide as the
region observed by Thomas et al.\ in AuCu under simular conditions. 

Wu et al.\ \cite{agcu-run} have studied the lattice misfit and misfit
dislocations in a silver on copper system with a 30 nm film of silver. After
irradiation they observe a lattice-mismatched interface below an epitaxial
layer, both in the copper substrate, covered by a 6 nm mixing region. The 2 nm
epitaxial Cu layer is nearly perfect bcc, while the mixing region shows a
transitional centered tetragonal structure. In their simulation the Ag and
part of the Cu substrate was molten. In the present simulations of Al on Ni no
lattice misfit was studied. The number of Ni to Al unit cells was optimized
before irradiation for minimal distortion leading to a fraction of
Ni:Al$=$27:31. After the irradiation the recrystallized structure is only 1 nm
thick while the Al film is still liquid and will turn amorphous if quenched.
There is no epitaxial layer although AlNi also solidifies into the bcc or B2
phase. However, there are lattice defects visible in the 8.5 and 9.5 nm slice
in Fig.~\ref{slices}. Since they are diagonal in the slices it is most likely
that they represent stacking defects in the fcc structure.

\subsection{Shock simulations}

The group of Strachnan
\cite{alni-shocksim,alni-amorph,alni-react,alni-last}
has intensively studied Al/Ni nanolaminates with different geometries and
composition, among them layered systems which are close to the Al films on Ni
presented here. They observe the exothermic reaction between Al and Ni as
expected. No additional heat production was observed here. The reason might be
that Strachnan et al. start with cold samples which are heated by shock wave
and the reaction, while laser irradiation adds a large amount of energy, such
that the additional reaction energy adds only a small effect. To find the
effect would thus required precise simulations with increasing fluences, which
have not yet been carried out.

\section{Summary and Conclusions}

Molecular dynamics simulations of laser melting and ablation of the AlNi
system have been reported. For pure metals experimental and simulation results
from other references could be reproduced. For the nickel aluminides 
reasonable results where obtained which largely confirm to the general formula
of Preuss et al. \cite{preuss1995-exp}. The interesting result in this context
is the changing ablation mechanism, in Al$_3$Ni accompanied by a phase
transition. Unfortunately, there are only very few experimental 
results available for these materials. Hopefully this study will encourage
experiments for these materials.

First results on coatings of Ni by Al have been obtained. There alloying of a
single layer system has been observed with a mixing gradient. Further studies
of parallel and orthogonal multi-layer systems would be very interesting. The
expected exothermal reaction of the nickel aluminate formation has not yet
been studied in detail since the present results did not indicate an
extraordinary behavior as expected from the shock wave studies.

Except for the weakly anisotropic orthorhombic Al$_3$Ni all the structures
studied here are isotropic due to the cubic symmetry. In
Ref.~\cite{itapdb:sonntag2009a} the MD+TTM model has been applied to strongly
anisotropic Al$_{13}$Co$_{4}$ with orthorhombic crystal 
structure. Then the heat conductivity has to be treated as a diagonal
tensor. However, it turned out that the ablation properties did only weakly
depend on the crystal direction. 

In conclusion, it was found that it is very involved to study complex
materials with the combined MD+TTM model. It is difficult to obtain all the
electronic and coupling parameters which for simplicity have been treated here
as constant. In general the parameters should be a function of the electron
temperature. The same is true for the atomic interactions
which should also be electron-temperature dependent. First attempts to
calculate such interactions exist, but only for pure metals.

From an experimental point of view there exists no obstacles to study other
materials and compounds which might be more relevant. 
From an industrial point of view there is great interest to treat for example 
steel instead of pure iron. The major advantage of the MD+TTM method is the
atomic resolution. Not only global observables like melting depths and
ablation thresholds can be studied but also the ablation mechanisms, phase
transformation, defect formation, and so on. Thus it is worth the effort.

\begin{theacknowledgments}
Financial funding from the German Science Foundation DFG for the Collaborative
Research Center SFB 716 ``Dynamic simulations of systems with large particle
numbers'' in subproject B.5 ``Radiation induced modification of charge state
and interaction by laser diffusion'' is greatly acknowledged.
\end{theacknowledgments}

%
% BibTeX users please use
% \bibliographystyle{}
% \bibliography{}

\begin{thebibliography}{9}
%Author, Journal \textbf{Volume,} (year) page numbers.
%Author, \textit{Book title} (Publisher, place year) page numbers
% etc

\bibitem{Schaefer2002} 
  C. Sch\"afer, H.M., Urbassek, L.V. Zhigilei,
  %  Metal ablation by picosecond laser pulses: A hybrid simulation,
  Phys. Rev. B \textbf{66}, 115404 (2002).

\bibitem{itapdb:sonntag2009} S. Sonntag, J. Roth, F. G\"ahler, H.-R. Trebin,
  %  Femtosecond laser ablation of aluminum.
  Appl. Surf. Sci. \textbf{255}, 9742 (2009).

\bibitem{itapdb:Roth2011}
  J. Roth, C. Trichet, H.-R. Trebin, S. Sonntag,
%  Laser ablation of metals,
in {\it High Performance Computing in Science and Engineering '10},
eds. W.E.Nagel, D.B. Kr\"oner, M.M. Resch, (Springer Heidelberg, 2011) pp. 159.

\bibitem{itapdb:Anisimov1974}
  S.I. Anisimov, B.L. Kapeliovich,
%  Electron-emission from surface of metals induced by ultrashort laser
%  pulses,   
  Perel'man T.L., Zh. Eksp. Teor. Fiz. \textbf{66}, 776 (1974)
  [Sov. Phys. JETP \textbf{39}, 375 (1974)].

\bibitem{itapdb:sonntag2009a} S. Sonntag, J. Roth, H.-R. Trebin,
% Molecular dynamics simulations of laser ablation in orthorhombic
% {A}l$_{13}${C}o$_4$.
  Appl. Phys. A \textbf{101}, 77 (2010).

\bibitem{aucu-ablat} D.A. Thomas, Z. Lin, L.V.Zhigilei, E.L. Gurevich,
  S. Kittel, R. Hergenr\"oder,
%  Atomistic modeling of femtosecond laser-induced melting and atomic mixing
%  in Au film - Cu substrate system,
Appl. Surf. Sci. \textbf{255}, 9605 (2009).

\bibitem{agcu-run}
  C. Wu, D.A. Thomas, Z. Lin, L.V. Zhigilei, 
%  Runaway lattice-mismatched interface in an atomistic simulation of
  %  femtosecond laser irradiation of Ag film - Cu substrate system,
  Appl. Phys. A \textbf {104}, 781 (2011).

\bibitem{exalcu}
  E.L. Gurevich, S. Kittel, R. Hergenr\"oder,
%  Experimental and numerical study of surface alloying by femtosecond laser
%  radiation,
  Appl. Surf. Sci. \textbf{258}, 2576 (2012).

\bibitem{aucr1} T.Q. Qiu, C.L. Tien, Int. J. Heat Mass Trans. \textbf{37} 2789
  (1994).

\bibitem{aucr2} T.Q. Qiu, T. Juhasz, C. Suarez, W.E. Bron, C.L. Tien,
  Int. J. Heat Mass Trans. \textbf{37} 2799
  
\bibitem{alni-shocksim} S. Zhao, T.C. Germann, A. Strachan,
% Atomistic simulations of shock-induced alloying reactions in Ni/Al
% nanolaminates, 
  J. Chem. Phys. \textbf{125}, 164707 (2006).

\bibitem{alni-laminate}
  D.P. Adams, M.A. Rodriguez, J.P. McDonald, M.M. Bai, E. Jones jr.,
  L. Brewer, J.J. Moore,
  % Reactive Ni/Ti nanolaminate,
  J. Appl. Phys. \textbf{106}, 093505 (2009).
  
\bibitem{alni-shocksim3} S. Zhao, T.C. Germann, A. Strachan,
%  Molecular dynamics simulation of dynamical response of perfect and porous
  %  Ni/Al nanolaminates under shock loading,
  Phys. Rev. B \textbf{76}, 014103 (2007). 

\bibitem{alni-shocksim4} S. Zhao, T.C. Germann, A. Strachan,
%  Melting and alloying of Ni/Al nanolaminates induced by shock loading: A
  %  molecular dynamics simulation study,
  Phys. Rev. B \textbf{76}, 014105 (2007).

\bibitem{alni-amorph}
  K.G. Vishnu, M.J. Cherukara, H. Kim, A. Strachan,
%  Amorphous Ni/Al nanoscale laminates as high-energy intermolecular reactive
  %  composites,
  Phys. Rev. B \textbf{85}, 184206 (2012).

\bibitem{alni-react} M.J. Cherukara, K.G. Vishnu, A. Strachan,
% Role of nanostructure on reaction and transport in Ni/Al intermolecular
  % reactive composites,
  Phys. Rev. B \textbf{86}, 075470 (2012).

\bibitem{alni-last} M.J. Cherukara, T.P. Weihs, A. Strachnan, Acta Mater
  \textbf{96} 1 (2015).

\bibitem{rothcola} J. Roth, H.-R. Trebin, A. Kiselev, D.-M- Rapp
  Appl. Phys. A \textbf{112} 500 (2016).
  
%\bibitem{ivanov2003}
%  D.S. Ivanov, L.V. Zhigilei,
%  Phys Rev. B \textbf{68}, 064114 (2003).

\bibitem{bunge-skript}
W.W. Baumann, U. Bunge, O. Frederich, M. Schatz, F. Thiele,
 \emph{Finite-Volumen-Methode in der Numerischen
  Thermofluiddynamik}.
 Berlin : Institut f\"ur Str{\"o}mungsmechanik und technische Akustik,
  TU Berlin, 2006
  (\url{http://www.cfd.tu-berlin.de/Lehre/CFD2/skript/fvm-skript.pdf}).

\bibitem{itapdb:Stadler1997a}
  J. Stadler, R. Mikulla, H.-R. Trebin,
% IMD: A software package for molecular dynamics studies on parallel
% computers.
  Int. J. Mod. Phys. C \textbf{8}, 1131 (1997).

\bibitem{itapdb:Roth2000} J. Roth, F. G\"ahler, H.-R. Trebin,
 % A molecular dynamics run with 5.180.116.000 particles,
  Int. J. Mod. Phys. C \textbf{11}, 317--322 (2000).

\bibitem{imd} The program is available at \url{http://itapmd.github.io/imd/}.

\bibitem{mishin2009-NiPot}
G.P. Purja~Pun, Y. Mishin,
% Development of an interatomic potential for the Ni-Al system.
Philosophical Magazine \textbf{89}, 3245 (2009).

\bibitem{alco} P. Brommer, F. G\"ahler, Phil. Mag. \textbf{86} 753 (2006).

\bibitem{potfit} P. Brommer, A. Kiselev, D. Schopf, P.  Beck, J. Roth,
  H.-R. Trebin, Mod. Sim. Mat. Sci. Eng. \textbf{23} 074002 (2015).
  
%\bibitem{baskes1992-EAM}
%M.I. Baskes,
%%  Modified embedded-atom potentials for cubic matrials and impurities.
%Phys. Rev. B \textbf{45}, 2727 (1992).

%\bibitem{bradley1937-NiAlXRay} A.J. Bradley, A. Taylor, W.L. Bragg,
%%  An X-Ray Analysis of the Nickel-Aluminum System.
%  Proc. R. Soc. Lond. A \textbf{159}, 56 (1937).
  
%\bibitem{sonntag2010}
%S. Sonntag, \emph{Computer Simulations of Laser Ablation from Simple Metals to
%   Complex Metallic Alloys}, Dissertation (Universit\"at Stuttgart 2010).

%\bibitem{baeuerle2000}
%D. B{\"a}uerle, \emph{Laser Processing and Chemistry}.
% (Springer-Verlag Berlin 2000).

%\bibitem{wellershoff1999-ElPh}
%S.S. Wellershoff, J. Hohlfeld, J. G\"udde, E. Matthias,
%% The role of electron--phonon coupling in femtosecond laser damage of metals.
%Appl. Phys. A \textbf{69}, 99 (1999).

%\bibitem{desai1987-ElHCP}
%P.D. Desai,
%% Thermodynamic properties of selected binary aluminium alloy systems.
%J. Phys. Chem. Ref. Data \textbf{16}, 109 (1987).

%\bibitem{terada2002-AlNiThCond}
%Y. Terada, K. Ohkubo, T. Mohri, T. Suzuki,
%% Thermal conductivity of intermetallic compounds with metallic bonding.
%Mater. Trans. \textbf{43}, 3167 (2002).

\bibitem{wang1994-ElPhHerleitung}
Z.G. Wang, C. Dufour, E. Paumier, M. Toulemonde,
% The $S_e$ sensitivity of metals under swift-heavy-ion irradiation: a
%  transient thermal process.
J. Phys., Condens. Matter \textbf{6}, 6733 (1994).

%\bibitem{stiehler2007-ElDens}
%M. Stiehler, U. Giegengack, J. Barzola-Quiquia, J. Rauchhaupt, S.
%Schulze, P. H\"aussler,
%% Peculiarities in the plasma resonance of binary amorphous Al-TM alloys.
%J. Phys. Chem. Solids \textbf{68}, 1244 (2007).

%\bibitem{stiehler2007-ElDens2}
%M. Stiehler, U. Giegengack, J. Rauchhaupt, P., H\"aussler,
% On modifications of the well-known Hume-Rothery rules: Amorphous alloys as
% model systems. 
%J. Non-Cryst. Solids \textbf{353}, 1886 (2007).

%\bibitem{saniz2006-OptAl3Ni}
%R. Saniz, L.-H. Ye, T. Shishidou, A.J. Freeman,
%% Structural, electronic and optical properties of NiAl$\mathsf{3}$:
%%  First-principles calculations.
%Phys. Rev. B \textbf{74}, 014209 (2006).

%\bibitem{rechtien1967-OptAlNi}
%J.J. Rechtien, C.R., Kannewurf, J.O., Brittain,
%% Optical Constants of $\beta$-Phase NiAl.
%J. Appl. Phys. \textbf{38}, 3045 (1967).

%\bibitem{hsu2004-OptAlNi3}
%  L.-S. Hsu, Y.-K., Wang,
%% Optical properties of Ni$\mathsf{_3}$Al, Ni$\mathsf{_3}$Ga and
%% Ni$\mathsf{_3}$In. 
%J. Alloy. Compd. \textbf{377}, 29 (2004).

%\bibitem{kandyla2007-AlMelt}
%M. Kandyla, T. Shih, E. Mazur,
%% Femtosecond dynamics of the laser-induced solid-to-liquid phase
%%  transition in aluminum.
%Phys. Rev. B \textbf{75}, 214107 (2007).

%\bibitem{guo2000-AlMelt}
%C. Guo, G. Rodriguez, A. Lobad, A.J., Taylor,
%% Structural Phase Transition of Aluminum Induced by Electronic
%%  Excitation.
%Phys. Rev. Lett. \textbf{84}, 4493 (2000).

%\bibitem{lewis2011-AlDamThs}
%M. Gill-Comeau, L.J., Lewis,
%% Ultrashort-pulse laser ablation of nanocrystalline aluminum.
%Phys. Rev. B \textbf{84}, 224110 (2011).

%\bibitem{povarnitsyn2007-AlHydro}
%M.E. Povarnitsyn, T.E. Itina, M. Sentis, K.V. Khishchenko, P.R. Levashov,
%% Material decomposition mechanisms in femtosecond laser interactions
%%  with metals.
%Phys. Rev. B \textbf{75}, 235414 (2007).

%\bibitem{amoruso2005-AlAbl}
%S. Amoruso, R. Bruzzese, M. Vitiello,
%% Experimental and theoretical investigations of femtosecond laser
%%  ablation of aluminum in vacuum.
%J. Appl. Phys. \textbf{98}, 044907 (2005).

%\bibitem{chimier2007-AlAbl}
%B. Chimier, V.T. Tikhonchuk, L. Hallo,
%% Heating model for metals irradiated by a subpicosecond laser pulse.
%Phys. Rev. B \textbf{75}, 195124 (2007).

%\bibitem{colombier2005-AlAbl}
%J.P. Colombier, P. Combis, F. Bonneau, R.L. Harzic, E. Audouard,
%% Hydrodynamic simulations of metal ablation by femtosecond laser
%%  irradiation.
%Phys. Rev. B \textbf{71}, 165406 (2005).

\bibitem{preuss1995-exp}
S. Preuss, A. Demchuk, M., Stuke,
% Sub-picosecond UV laser ablation of metals.
Appl. Phys. A \textbf{61}, 33 (1995).

%\bibitem{xu2002-NiMelt}
%D.A. Willis, X. Xu,
%% Heat transfer and phase change during picosecond laser ablation of nickel.
%Int. J. Heat. Mass. Transf. \textbf{45}, 3911 (2002).

%\bibitem{zhigi2007-ElPh}
%Z. Lin, L.V. Zhigilei,
%% Temperature dependences of the electron--phonon coupling, electron
%%  heat capacity and thermal conductivity in Ni under femtosecond laser
%%  irradiation.
%Appl. Surf. Sci. \textbf{253}, 6295 (2007).

%\bibitem{amoruso2007-NiAbl}
%S. Amoruso, R. Bruzzese, X. Wang, N.N. Nedialkov, P.A. Atanasov,
%% Femtosecond laser ablation of nickel in vacuum.
%J. Phys. D: Appl. Phys. \textbf{40}, 331 (2007).

%\bibitem{cheng2005-Mechanismen}
%C. Cheng, X. Xu,
%% Mechanisms of decomposition of metal during femtosecond laser
%%  ablation.
%Phys. Rev. B \textbf{72}, 165415 (2005).

%\bibitem{preuss1994-exp}
%S. Preuss, E. Matthias, M. Stuke,
%% Sub-picosecond UV laser ablation of Ni films: Strong fluence
%%  reduction and thickness-independent removal.
%Appl. Phys. A \textbf{59}, 79 (1994).

%\bibitem{atanasov2002-NiMD}
%P.A. Atanasov, N.N. Nedialkov, S.E. Imamova, A. Ruf, H. H\"ugel, F. Dausinger,
%P. Berger, 
%% Laser ablation of Ni by ultrashort pulses: molecular dynamics
%%  simulation.
%Appl. Surf. Sci. \textbf{186}, 369 (2002).

\bibitem{pollock2007-AlNi3Abl}
S. Ma, J.P. McDonald, B. Tryon, S.M. Yalisove, T.M. Pollock,
% Femtosecond laser ablation regimes in a single-crystal superalloy.
Metall. Mater. Trans. A \textbf{38A}, 2349 (2007).

\bibitem{guedde1998-NiDamThs}
J. G\"udde, J. Hohlfeld, J.G. M\"uller, E. Matthias,
% Damage threshold dependence on electron-phonon coupling in Au and Ni films.
Appl. Surf. Sci. \textbf{127-129}, 40 (1998).

%\bibitem{vanderheide1985-OptAlNi3}
%P.A.M. v.d.Heide, J.J.M. Buiting, L.M. Dam, L.W.M. Schreus, R.A. Groot,
%A.R. Vroomen,
%% Spectroscopic ellipsometry of Ni$\mathsf{_3}$Al in comparison with
%%  band-structure calculations.
%J. Phys. F: Met. Phys. \textbf{15}, 1195 (1985).

%\bibitem{yu}
% P. Yu, C.-J. Deng, N.-G. Ma, D.H.L. Ng,
%J. Mater. Res. \textbf{19} ,1187 (2004).
  
%\bibitem{robertson}
% I.M. Robertson, C.M., Wyman,
%  Metallography \textbf{17}, 43 (1984).

\bibitem{ovito}
  A. Stukowski,
  Model. Simul. Mater. Sci. \textbf{18}, 015012 (2010).

\end{thebibliography}
\bibliographystyle{aipproc}

%%%%%%%%%%%%%%%%%%%%%%%%%%%%%%%%%%%%%%%%%%%%%%%%%%%%%%%%%%%%%%%%%%%%%%
%
%%%%%%%%%%%%%%%%%%%%%%%%%%%%%%%%%%%%%%%%%%%%%%%%%%%%%%%%%%%%%%%%%%%%%%

%\printindex
\end{document}